\documentclass{ws-ijmpb}
\begin{document}

\markboth{Jian-Guo Liu, et. al}{Optimization of scale-free network
for random failures}

%
\catchline{}{}{}{}{}
%

\title{OPTIMIZATION OF SCALE-FREE NETWORK FOR RANDOM
FAILURES
}

\author{JIAN-GUO LIU$^\dag$, ZHONG-TUO WANG and YAN-ZHONG
DANG}

\address{Institute of System Engineering, Dalian University of
Technology, 2 Ling Gong Rd.,\\ Dalian 116024, P R China\\
$^\dag$liujg004@tom.com}

\maketitle

\begin{history}
\received{(8 February 2005)} \revised{(31 Aug 2005)}
\end{history}

\begin{abstract}
It has been found that the networks with scale-free distribution are
very resilient to random failures. The purpose of this work is to
determine the network design guideline which maximize the network
robustness to random failures with the average number of links per
node of the network is constant. The optimal value of the
distribution exponent and the minimum connectivity to different
network size are given in this paper. Finally, the optimization
strategy how to improve the evolving network robustness is given.
\end{abstract}


\section{Introduction}

Recently, much attention has been focused on the topic of
scale-free networks which characterize many social, information,
technological and biological systems.\cite{1,2,3,4} The
qualitative properties of many interesting real-world examples,
such as the internet network, the power grid network and the
protein interaction network, are as following:
\begin{description}
\vspace{4pt} \item[{(1)}] the degree distribution has power-law
tail; \vspace{4pt}\item[{(2)}] local clustering of edges: graph is
not locally tree-like; \vspace{4pt}\item[{(3)}] small average
distance.
\end{description}
The networks can be visualized by nodes representing individuals,
organizations, computers and by links between them representing
their interactions. For the purpose of analyzing topology, we
ignore the variation in the type of links. Robustness of the
network topology comes from the presence of alternate paths, which
ensures the communication remains possible in spite of the damages
to the network.

Designers of the networks must assume that networks have random
failures or might be attacked, and some of these attacks can
result in damage. The robust networks will continue functioning in
spite of such damages. Although many literatures have discussed
what the optimal network topology would be, many real-world
networks present the power-law degree distribution.

When the scale-free networks are subjected to random breakdowns,
with a fraction $p$ of the nodes and their connections are removed
randomly, the network's integrity might be compromised: when the
exponent of the power-law degree distribution $\gamma>3$, there
exists a critical threshold $p_c$, such that for $p>p_c$, the
network would break into smaller and disconnected parts, but the
networks with $\gamma<3$ are more resilient to random breakdowns.
Cohen {\textit {et. al}}\cite{10} presented a criterion to calculate
the percolation critical threshold to random failures to scale-free
networks. If we attack the scale-free networks intentionally: the
removal of sites is not random, but rather sites with the highest
connectivity are targeted first. The numerical simulations suggest
that scale-free networks are highly sensitive to this kind of
attack.\cite{9} Cohen {\textit {et. al}}\cite{11} studied the exact
value of the critical fraction needed for disruption. Thus
scale-free networks are highly robust against random failures of
nodes and hypersensitive to intentional attacks against the system's
largest nodes. So a randomly chosen node has low degree with high
probability, but removal of a highly connected node might produce
large effect to the network. This situation is often compared to
that of the classical random graph presented by Erd$\check{o}$s and
R\'{e}nyi.\cite{5,6} Such graphs have a Poisson degree distribution.
This makes the random graphs less robust to random failures than
comparable networks with power-law degree distribution, but much
more robust against attacks on hubs.


In this paper, we specifically focus on the robustness of the
network topology to random failures. We use the percolation theory
and the optimization method to investigate the guideline which can
maximize the robustness of the scale-free networks to random
failures of nodes with the constrained condition that the average
connectivity of per node in the network is constant. The percolation
theory provides the measures of distribution which are possible ways
for measuring robustness. We examine the relationship between the
average connectivity per node and the network robustness to random
failures. Then, we investigate the trend of the network robustness
to random failures with the network size $N$. The work may provide
the theoretical evidence that if the minimal connectivity and the
exponent of the power-law degree distribution take in more optimal
way, the robustness of the scale-free networks can be optimized.

If we construct and maintain a network with a given number of
nodes as being proportional to the average number of links
$\langle k \rangle$ per node in the network, our goal then becomes
how to maximize the robustness of a network with $N$ nodes to
random failures with the constraint that the number of links
remains constant but the nodes are connected in a different and
more optimal way.

\section{Optimal Strategy for Random Failures}

Our goal is to maximize the threshold for random removal with the
condition that the average degree $\langle k \rangle$ per node is
constant. We construct the following model.
\begin{equation}\label{F2.7}
  \left\{\begin{array}{rlc}
    \max & p_c^{\rm rand}, \\[5pt]
    {\rm s.t.} & \langle k \rangle={\rm constant.}
  \end{array}
  \right.
\end{equation}

For any degree distribution $P(k)$, the threshold for random
removal of nodes is$^{9}$

\begin{equation}\label{F2.1}
p_c^{\rm rand}=1-\frac{1}{\kappa_0-1},
\end{equation}
where $\kappa_0\equiv \frac{\langle k^2\rangle}{\langle k
\rangle}$ is calculated from the original connectivity
distribution. A wide range of networks have the power-law degree
distribution:
\begin{equation}\label{F2.2}
P(k)=ck^{-\alpha}, \ \ k=m, m+1, \ldots, K,
\end{equation}
where $k=m$ is the minimal connectivity and $k=K$ is an effective
connectivity cutoff presented in finite networks. To the power-law
degree distribution, the average $\langle k \rangle$ can be given
with the usual continuous approximation, this yields
\begin{equation}\label{F2.3}
 \langle k \rangle = \int_m^K{k\cdot ck^{-\alpha}}dk
     =  c\frac{[K^{(2-\alpha)}-m^{(2-\alpha)}]}{2-\alpha}.
\end{equation}
 From (\ref{F2.2}), $\kappa_0$ can be calculated as
\begin{equation}\label{F2.3}
  \kappa_0  =  \langle k^2\rangle/\langle k \rangle
       =
      \frac{2-\alpha}{3-\alpha}\frac{[K^{(2-\alpha)}-m^{(2-\alpha)}]}{[K^{(3-\alpha)}-m^{(3-\alpha)}]}.
\end{equation}
In a finite network, the largest connectivity $K$ can be estimated
from\cite{10}
$$
\int_K^{\infty}P(k)dk=\frac{1}{N},
$$
where $N$ is the number of the network nodes. Then we have that
\begin{equation}\label{F2.6}
\big(\frac{K}{m}\big)^{\alpha-1}=N.
\end{equation}
To the power-law degree distribution $P(k)$, we have
$$
\begin{array}{rcl}
    1 & = & \int_m^K ck^{-\alpha}dk \\[8pt]
      & = & c[K^{(1-\alpha)}-m^{(1-\alpha)}]/(1-\alpha), \\[8pt]
\end{array}
$$
this yields
$$
  c=\frac{\alpha-1}{m^{(1-\alpha)}-K^{(1-\alpha)}}=\frac{m^{\alpha}(\alpha-1)}{m[1-(\frac{K}{m})^{(1-\alpha)}]}.
$$
In the real world, there always exists the relation $K\gg m$, so
we have
\begin{equation}\label{F2.4}
c\approx \frac{m^{\alpha}(\alpha-1)}{m}.
\end{equation}
Combining (\ref{F2.3}) and (\ref{F2.4}), we have that
\begin{equation}\label{F2.5}
\langle k \rangle\approx
\frac{(\alpha-1)}{(2-\alpha)}m\big[\big(\frac{K}{m}\big)^{(2-\alpha)}-1\big].
\end{equation}
From (\ref{F2.5}), we have the following numerical results.

\vspace{0.3cm}
 It can been seen from table 1 that the
 distribution exponent $\alpha$ increases when the minimal connectivity $m$ increases.

Combining (\ref{F2.6}) and (\ref{F2.5}), we have that
\begin{equation}\label{F2.15}
\langle k \rangle=
\frac{(\alpha-1)}{(\alpha-2)}m\frac{1-N^{-\frac{\alpha-2}{\alpha-1}}}{1-N^{-1}}.
\end{equation}
\begin{table}[pt]
\tbl{When $N=10^6$, the relationship between $m$ and $p_c^{\rm
rand}$.} {\begin{tabular}{cccccccc@{}} \toprule
   $\langle k\rangle$       & $a(m=1)$  & $a(m=2)$    & $a(m=3)$   &  $a(m=4)$     &  $a(m=5)$ & $a(m=6)$\\
\colrule
     3 & 2.492    &  4.000     & --       &  --      &  --    & --    \\
     4 & 2.318    &  2.998     & 5.000    &  --      &  --    & --    \\
     5 & 2.225    &  2.662     & 3.498    &  6.000   &  --    & --    \\
     6 & 2.172    &  2.491     & 2.998    &  4.000   &  7.00  & --    \\
     7 & 2.126    &  2.388     & 2.747    &  3.333   &  4.50  & 8.00  \\ \botrule
\end{tabular} }
\end{table}

From table 1, we can get the following relationship:
\begin{description}
\item [(1)] When the average connectivity $\langle k \rangle$ per
node is constant, the exponent $\alpha$ increases when the minimum
connectivity $m$ increases;

\item[(2)] To the minimum connectivity $m=1$, the exponent
$\alpha$ decreases when the average connectivity $\langle k
\rangle$ of the network increases.
\end{description}

Using the results obtained above we construct the following model.
\begin{equation}\label{F2.8}
  \left\{\begin{array}{rl}
    \max & \{1-\frac{1}{\kappa_0-1}\}
    \\[10pt]
     s.t.
    &\frac{(\alpha-1)}{(\alpha-2)}m[1-N^{-\frac{\alpha-2}{\alpha-1}}]=\langle k \rangle\\[10pt]
    & m\in Z^{+},
  \end{array}
  \right.
\end{equation}
where
$\kappa_0=\frac{2-\alpha}{3-\alpha}\frac{[K^{(2-\alpha)}-m^{(2-\alpha)}]}{[K^{(3-\alpha)}-m^{(3-\alpha)}]}$.
The numerical results suggest that whether the network size $N$ is
very large or not, $p_c$ reaches its maximum value when $m=1$. The
numerical results are presented in table 2.


\begin{figure}[th]
\centerline{\psfig{file=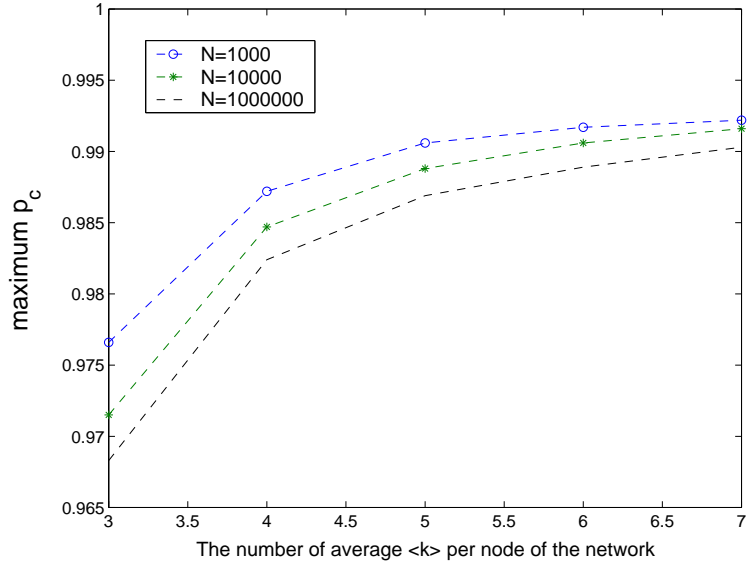,width=10cm}} \vspace*{8pt}
\caption{The critical percolation thresholds for random failures
to different $\langle k \rangle$.}
\end{figure}

\begin{table}[pt]
\tbl{When $N=10^6$, the relationship between $m$ and $p_c^{\rm
rand}$.} {\begin{tabular}{cccccccc@{}} \toprule
              & $\langle k \rangle$=3   &  $\langle k \rangle$=4     & $\langle k \rangle$=5    &  $\langle k \rangle$=6   &  $\langle k \rangle$=7 \\
\colrule
     $N=10^3$ & 0.9766    &  0.9872      & 0.9906     &  0.9917    &  0.9922  \\
     $N=10^4$ & 0.9715    &  0.9847      & 0.9888     &  0.9906    &  0.9916  \\
     $N=10^6$ & 0.9683    &  0.9824      & 0.9869     &  0.9889    &  0.9903  \\ \botrule
\end{tabular} }
\end{table}

From table 2, we can get the following three conclusions:
\begin{description}
\item[{(1)}] If the average connectivity $\langle k \rangle$ per
node and the exponent $\alpha$ of the power-law degree
distribution is constant, the robustness of the scale-free
networks will decrease when the network size becomes larger.

\item[{(2)}] If the network size $N$ is constant, the robustness
of the scale-free networks increase when the average connectivity
$\langle k \rangle$ becomes larger.

\item[{(3)}]  To the random failures, we have to take several
times cost to increase the robustness of the scale-free networks
about 1\%.
\end{description}

\section{Discussion and Summary}

 It is well known that the networks with power-law degree distribution
 are resilient to random failures. But this conclusion don't answer the
following three questions: (i) To a constant average connectivity
$\langle k \rangle$, how to determine the distribution exponent of
the scale-free networks so that the networks are more robust to the
random failures. (ii) To different network size, how many edges we
need to add to the network to satisfy the robustness level to random
failures.  (iii) To an exist network with power-law degree
distribution, what we need to do to improve the network robustness.
In this paper, we use the percolation theory and the mathematic
programme method to optimize the robustness of the scale-free
networks for random failures and give the numerical results.
Finally, we give the relationship between the threshold $p_c$ and
the network size, the degree distribution exponent $\alpha$ and the
average connectivity $\langle k \rangle$ per node.

From Fig. 1, we can get the conclusion that if the scale-free
networks size become large, the network robustness to random
failures would become weak. To the internet and other growing
scale-free networks the designers must add more links to the network
to improve the average connectivity per node to random failures.

Subjects for further study include (i) an analysis of the
robustness of the scale-free networks to the intentional attack to
the highest connectivity nodes. (ii) the optimization of complex
network under both random failures and intentional attack. (iii)
the topology structure to improving the robustness of existing
scale-free networks.

\section*{Acknowledgments}
The authors are grateful to Dr. Qiang Guo for her valuable
comments and suggestions, which have led to a better presentation
of this paper. This research was supported by Chinese Natural
Science Foundation Grant Nos. 70431001 and 70271046.



\begin{thebibliography}{0}
\bibitem{1} R. Albert and A.-L. Barab\'{a}si, {\it Rev. Mod. Phys.} {\bf
74}, 47 (2002).

\bibitem{2} M. E. J. Newmann, {\it SIAM Rev.} {\bf 45}, 167 (2003).

\bibitem{3} J. F. F. Mendes, S. N. Dorogovtsev and A. F. Ioffe, {\it
Evolution of Networks: From Biological Nets to the Internet and
the WWW} (Oxford University Press, Oxford, 2003).

\bibitem{4} R. Pastor-Satorras and A. Vespignani, {\it Evolution and
Structure of the Internet: A Statistical Physics Approach}
(Cambridge University Press, Cambridge, 2004).

\bibitem{10} R. Cohen, K, Erez, D. ben-Avraham, S. Havlin, {\it Phys.
Rev. Lett.} {\bf 85}, 4626 (2000).

\bibitem{9} G. Paul, T. Tanizawa, S. Havlin and H. E. Stanley, {\it Eur.
Phys. J. B} {\bf 38}, 187 (2004).

\bibitem{11} R. Cohen, K, Erez, D. ben-Avraham, S. Havlin, {\it Phys.
Rev. Lett.} {\bf 86}, 3682 (2001).



\bibitem{5} P. Erd\u{o}s and  A. R\'{e}nyi, {\it Publicationes
Mathematicae} {\bf 6}, 290 (1959).


\bibitem{6} P. Erd\u{o}s and  A. R\'{e}nyi, {\it Publicationes
Mathematical Inst. of the Hungarian Acad. of Sciences} {\bf 5}, 17
(1960).

\bibitem{7} D. J. Watts and S. H. Strogatz, {\it Nature} {\bf 393}, 440
(1998).

\bibitem{8} B. Bollobas, {\it Random Graphs} (Academic, London, 1985).


\bibitem{12} R. Albert, H. Jeong, and A. L. Barab\'{a}si, {\it Nature},
{\bf 406}, 6794 (2000).


\end{thebibliography}
\end{document}